%% file: main.tex
\documentclass[a4paper,11pt]{article}
\usepackage{pos}
\usepackage{units}

\title{Science and mission status of EUSO-SPB2}
 \ShortTitle{EUSO-SPB2}

\author*[a]{J. Eser}
\author[a]{A. V. Olinto}
\author[b]{L. Wiencke}

\affiliation[a]{Department of Astronomy \& Astrophysics, University of Chicago,\\
Chicago, IL 60637, USA}

\affiliation[b]{Department of Physics, Colorado School of Mines,\\
Golden, CO 80401, USA}

\forColl{JEM-EUSO} 

\emailAdd{jeser@uchicago.edu}

\abstract{The Extreme Universe Space Observatory on a Super Pressure Balloon II (EUSO-SPB2) is a second generation stratospheric balloon instrument for the detection of Ultra High Energy Cosmic Rays (UHECRs, E > 1~EeV) via the fluorescence technique and of Very High Energy (VHE, E > 10~PeV) neutrinos via Cherenkov emission. EUSO-SPB2 is a pathfinder mission for instruments like the proposed Probe Of Extreme Multi-Messenger Astrophysics (POEMMA). The purpose of such a space-based observatory is to measure UHECRs and UHE neutrinos with high statistics and uniform exposure.\\
EUSO-SPB2 is designed with two Schmidt telescopes, each optimized for their respective observational goals. The Fluorescence Telescope looks at the nadir to measure the fluorescence emission from UHECR-induced extensive air shower (EAS), while the Cherenkov Telescope is optimized for fast signals ($\sim$10~ns) and points near the Earth's limb. This allows for the measurement of Cherenkov light from EAS caused by Earth skimming VHE neutrinos if pointed slightly below the limb or from UHECRs if observing slightly above.\\
The expected launch date of EUSO-SPB2 is Spring 2023 from Wanaka, NZ with target duration of up to 100 days. Such a flight would provide thousands of VHECR Cherenkov signals in addition to tens of UHECR fluorescence tracks. Neither of these kinds of events have been observed from either orbital or suborbital altitudes before, making EUSO-SPB2 crucial to move forward towards a space-based instrument. It will also enhance the understanding of potential background signals for both detection techniques.\\
This contribution will provide a short overview of the detector and the current status of the mission as well as its scientific goals.}

\FullConference{37$^{\rm{th}}$ International Cosmic Ray Conference (ICRC 2021)\\
		July 12th -- 23rd, 2021\\
		Online -- Berlin, Germany}

\begin{document}
\maketitle

\input{intro}
\input{Mission_Instrument}
\input{FT_science}
\input{CT_science}
\input{Conclusion}

\bibliographystyle{JHEP}
\bibliography{ICRC2021ref.bib}

\clearpage
\section*{Full Authors List: \Coll\ Collaboration}
%
\begin{sloppypar}
\scriptsize
\noindent
G.~Abdellaoui$^{ah}$, 
S.~Abe$^{fq}$, 
J.H.~Adams Jr.$^{pd}$, 
D.~Allard$^{cb}$, 
G.~Alonso$^{md}$, 
L.~Anchordoqui$^{pe}$,
A.~Anzalone$^{eh,ed}$, 
E.~Arnone$^{ek,el}$,
K.~Asano$^{fe}$,
R.~Attallah$^{ac}$, 
H.~Attoui$^{aa}$, 
M.~Ave~Pernas$^{mc}$,
M.~Bagheri$^{ph}$,
J.~Bal\'az{$^{la}$, 
M.~Bakiri$^{aa}$, 
D.~Barghini$^{el,ek}$,
S.~Bartocci$^{ei,ej}$,
M.~Battisti$^{ek,el}$,
J.~Bayer$^{dd}$, 
B.~Beldjilali$^{ah}$, 
T.~Belenguer$^{mb}$,
N.~Belkhalfa$^{aa}$, 
R.~Bellotti$^{ea,eb}$, 
A.A.~Belov$^{kb}$, 
K.~Benmessai$^{aa}$, 
M.~Bertaina$^{ek,el}$,
P.F.~Bertone$^{pf}$,
P.L.~Biermann$^{db}$,
F.~Bisconti$^{el,ek}$, 
C.~Blaksley$^{ft}$, 
N.~Blanc$^{oa}$,
S.~Blin-Bondil$^{ca,cb}$, 
P.~Bobik$^{la}$, 
M.~Bogomilov$^{ba}$,
E.~Bozzo$^{ob}$,
S.~Briz$^{pb}$, 
A.~Bruno$^{eh,ed}$, 
K.S.~Caballero$^{hd}$,
F.~Cafagna$^{ea}$, 
G.~Cambi\'e$^{ei,ej}$,
D.~Campana$^{ef}$, 
J-N.~Capdevielle$^{cb}$, 
F.~Capel$^{de}$, 
A.~Caramete$^{ja}$, 
L.~Caramete$^{ja}$, 
P.~Carlson$^{na}$, 
R.~Caruso$^{ec,ed}$, 
M.~Casolino$^{ft,ei}$,
C.~Cassardo$^{ek,el}$, 
A.~Castellina$^{ek,em}$,
O.~Catalano$^{eh,ed}$, 
A.~Cellino$^{ek,em}$,
K.~\v{C}ern\'{y}$^{bb}$,  
M.~Chikawa$^{fc}$, 
G.~Chiritoi$^{ja}$, 
M.J.~Christl$^{pf}$, 
R.~Colalillo$^{ef,eg}$,
L.~Conti$^{en,ei}$, 
G.~Cotto$^{ek,el}$, 
H.J.~Crawford$^{pa}$, 
R.~Cremonini$^{el}$,
A.~Creusot$^{cb}$, 
A.~de Castro G\'onzalez$^{pb}$,  
C.~de la Taille$^{ca}$, 
L.~del Peral$^{mc}$, 
A.~Diaz Damian$^{cc}$,
R.~Diesing$^{pb}$,
P.~Dinaucourt$^{ca}$,
A.~Djakonow$^{ia}$, 
T.~Djemil$^{ac}$, 
A.~Ebersoldt$^{db}$,
T.~Ebisuzaki$^{ft}$,
L.~Eliasson$^{na}$, 
J.~Eser$^{pb}$,
F.~Fenu$^{ek,el}$, 
S.~Fern\'andez-Gonz\'alez$^{ma}$, 
S.~Ferrarese$^{ek,el}$,
G.~Filippatos$^{pc}$, 
 W.I.~Finch$^{pc}$
C.~Fornaro$^{en,ei}$,
M.~Fouka$^{ab}$, 
A.~Franceschi$^{ee}$, 
S.~Franchini$^{md}$, 
C.~Fuglesang$^{na}$, 
T.~Fujii$^{fg}$, 
M.~Fukushima$^{fe}$, 
P.~Galeotti$^{ek,el}$, 
E.~Garc\'ia-Ortega$^{ma}$, 
D.~Gardiol$^{ek,em}$,
G.K.~Garipov$^{kb}$, 
E.~Gasc\'on$^{ma}$, 
E.~Gazda$^{ph}$, 
J.~Genci$^{lb}$, 
A.~Golzio$^{ek,el}$,
C.~Gonz\'alez~Alvarado$^{mb}$, 
P.~Gorodetzky$^{ft}$, 
A.~Green$^{pc}$,  
F.~Guarino$^{ef,eg}$, 
C.~Gu\'epin$^{pl}$,
A.~Guzm\'an$^{dd}$, 
Y.~Hachisu$^{ft}$,
A.~Haungs$^{db}$,
J.~Hern\'andez Carretero$^{mc}$,
L.~Hulett$^{pc}$,  
D.~Ikeda$^{fe}$, 
N.~Inoue$^{fn}$, 
S.~Inoue$^{ft}$,
F.~Isgr\`o$^{ef,eg}$, 
Y.~Itow$^{fk}$, 
T.~Jammer$^{dc}$, 
S.~Jeong$^{gb}$, 
E.~Joven$^{me}$, 
E.G.~Judd$^{pa}$,
J.~Jochum$^{dc}$, 
F.~Kajino$^{ff}$, 
T.~Kajino$^{fi}$,
S.~Kalli$^{af}$, 
I.~Kaneko$^{ft}$, 
Y.~Karadzhov$^{ba}$, 
M.~Kasztelan$^{ia}$, 
K.~Katahira$^{ft}$, 
K.~Kawai$^{ft}$, 
Y.~Kawasaki$^{ft}$,  
A.~Kedadra$^{aa}$, 
H.~Khales$^{aa}$, 
B.A.~Khrenov$^{kb}$, 
 Jeong-Sook~Kim$^{ga}$, 
Soon-Wook~Kim$^{ga}$, 
M.~Kleifges$^{db}$,
P.A.~Klimov$^{kb}$,
D.~Kolev$^{ba}$, 
I.~Kreykenbohm$^{da}$, 
J.F.~Krizmanic$^{pf,pk}$, 
K.~Kr\'olik$^{ia}$,
V.~Kungel$^{pc}$,  
Y.~Kurihara$^{fs}$, 
A.~Kusenko$^{fr,pe}$, 
E.~Kuznetsov$^{pd}$, 
H.~Lahmar$^{aa}$, 
F.~Lakhdari$^{ag}$,
J.~Licandro$^{me}$, 
L.~L\'opez~Campano$^{ma}$, 
F.~L\'opez~Mart\'inez$^{pb}$, 
S.~Mackovjak$^{la}$, 
M.~Mahdi$^{aa}$, 
D.~Mand\'{a}t$^{bc}$,
M.~Manfrin$^{ek,el}$,
L.~Marcelli$^{ei}$, 
J.L.~Marcos$^{ma}$,
W.~Marsza{\l}$^{ia}$, 
Y.~Mart\'in$^{me}$, 
O.~Martinez$^{hc}$, 
K.~Mase$^{fa}$, 
R.~Matev$^{ba}$, 
J.N.~Matthews$^{pg}$, 
N.~Mebarki$^{ad}$, 
G.~Medina-Tanco$^{ha}$, 
A.~Menshikov$^{db}$,
A.~Merino$^{ma}$, 
M.~Mese$^{ef,eg}$, 
J.~Meseguer$^{md}$, 
S.S.~Meyer$^{pb}$,
J.~Mimouni$^{ad}$, 
H.~Miyamoto$^{ek,el}$, 
Y.~Mizumoto$^{fi}$,
A.~Monaco$^{ea,eb}$, 
J.A.~Morales de los R\'ios$^{mc}$,
M.~Mastafa$^{pd}$, 
S.~Nagataki$^{ft}$, 
S.~Naitamor$^{ab}$, 
T.~Napolitano$^{ee}$,
J.~M.~Nachtman$^{pi}$
A.~Neronov$^{ob,cb}$, 
K.~Nomoto$^{fr}$, 
T.~Nonaka$^{fe}$, 
T.~Ogawa$^{ft}$, 
S.~Ogio$^{fl}$, 
H.~Ohmori$^{ft}$, 
A.V.~Olinto$^{pb}$,
Y.~Onel$^{pi}$
G.~Osteria$^{ef}$,  
A.N.~Otte$^{ph}$,  
A.~Pagliaro$^{eh,ed}$, 
W.~Painter$^{db}$,
M.I.~Panasyuk$^{kb}$, 
B.~Panico$^{ef}$,  
E.~Parizot$^{cb}$, 
I.H.~Park$^{gb}$, 
B.~Pastircak$^{la}$, 
T.~Paul$^{pe}$,
M.~Pech$^{bb}$, 
I.~P\'erez-Grande$^{md}$, 
F.~Perfetto$^{ef}$,  
T.~Peter$^{oc}$,
P.~Picozza$^{ei,ej,ft}$, 
S.~Pindado$^{md}$, 
L.W.~Piotrowski$^{ib}$,
S.~Piraino$^{dd}$, 
Z.~Plebaniak$^{ek,el,ia}$, 
A.~Pollini$^{oa}$,
E.M.~Popescu$^{ja}$, 
R.~Prevete$^{ef,eg}$,
G.~Pr\'ev\^ot$^{cb}$,
H.~Prieto$^{mc}$, 
M.~Przybylak$^{ia}$, 
G.~Puehlhofer$^{dd}$, 
M.~Putis$^{la}$,   
P.~Reardon$^{pd}$, 
M.H..~Reno$^{pi}$, 
M.~Reyes$^{me}$,
M.~Ricci$^{ee}$, 
M.D.~Rodr\'iguez~Fr\'ias$^{mc}$, 
O.F.~Romero~Matamala$^{ph}$,  
F.~Ronga$^{ee}$, 
M.D.~Sabau$^{mb}$, 
G.~Sacc\'a$^{ec,ed}$, 
G.~S\'aez~Cano$^{mc}$, 
H.~Sagawa$^{fe}$, 
Z.~Sahnoune$^{ab}$, 
A.~Saito$^{fg}$, 
N.~Sakaki$^{ft}$, 
H.~Salazar$^{hc}$, 
J.C.~Sanchez~Balanzar$^{ha}$,
J.L.~S\'anchez$^{ma}$, 
A.~Santangelo$^{dd}$, 
A.~Sanz-Andr\'es$^{md}$, 
M.~Sanz~Palomino$^{mb}$, 
O.A.~Saprykin$^{kc}$,
F.~Sarazin$^{pc}$,
M.~Sato$^{fo}$, 
A.~Scagliola$^{ea,eb}$, 
T.~Schanz$^{dd}$, 
H.~Schieler$^{db}$,
P.~Schov\'{a}nek$^{bc}$,
V.~Scotti$^{ef,eg}$,
M.~Serra$^{me}$, 
S.A.~Sharakin$^{kb}$,
H.M.~Shimizu$^{fj}$, 
K.~Shinozaki$^{ia}$, 
T.~Shirahama$^{fn}$,
J.F.~Soriano$^{pe}$,
A.~Sotgiu$^{ei,ej}$,
I.~Stan$^{ja}$, 
I.~Strharsk\'y$^{la}$, 
N.~Sugiyama$^{fj}$, 
D.~Supanitsky$^{ha}$, 
M.~Suzuki$^{fm}$, 
J.~Szabelski$^{ia}$,
N.~Tajima$^{ft}$, 
T.~Tajima$^{ft}$,
Y.~Takahashi$^{fo}$, 
M.~Takeda$^{fe}$, 
Y.~Takizawa$^{ft}$, 
M.C.~Talai$^{ac}$, 
Y.~Tameda$^{fu}$, 
C.~Tenzer$^{dd}$,
S.B.~Thomas$^{pg}$, 
O.~Tibolla$^{he}$,
L.G.~Tkachev$^{ka}$,
T.~Tomida$^{fh}$, 
N.~Tone$^{ft}$, 
S.~Toscano$^{ob}$, 
M.~Tra\"{i}che$^{aa}$, 
Y.~Tsunesada$^{fl}$, 
K.~Tsuno$^{ft}$,  
S.~Turriziani$^{ft}$, 
Y.~Uchihori$^{fb}$, 
O.~Vaduvescu$^{me}$, 
J.F.~Vald\'es-Galicia$^{ha}$, 
P.~Vallania$^{ek,em}$,
L.~Valore$^{ef,eg}$,
G.~Vankova-Kirilova$^{ba}$, 
T.~M.~Venters$^{pj}$,
C.~Vigorito$^{ek,el}$, 
L.~Villase\~{n}or$^{hb}$,
B.~Vlcek$^{mc}$, 
P.~von Ballmoos$^{cc}$,
M.~Vrabel$^{lb}$, 
S.~Wada$^{ft}$, 
J.~Watanabe$^{fi}$, 
J.~Watts~Jr.$^{pd}$, 
R.~Weigand Mu\~{n}oz$^{ma}$, 
A.~Weindl$^{db}$,
L.~Wiencke$^{pc}$, 
M.~Wille$^{da}$, 
J.~Wilms$^{da}$,
D.~Winn$^{pm}$
T.~Yamamoto$^{ff}$,
J.~Yang$^{gb}$,
H.~Yano$^{fm}$,
I.V.~Yashin$^{kb}$,
D.~Yonetoku$^{fd}$, 
S.~Yoshida$^{fa}$, 
R.~Young$^{pf}$,
I.S~Zgura$^{ja}$, 
M.Yu.~Zotov$^{kb}$,
A.~Zuccaro~Marchi$^{ft}$
}
\end{sloppypar}

{ \footnotesize
\noindent
$^{aa}$ Centre for Development of Advanced Technologies (CDTA), Algiers, Algeria \\
$^{ab}$ Dep. Astronomy, Centre Res. Astronomy, Astrophysics and Geophysics (CRAAG), Algiers, Algeria \\
$^{ac}$ LPR at Dept. of Physics, Faculty of Sciences, University Badji Mokhtar, Annaba, Algeria \\
$^{ad}$ Lab. of Math. and Sub-Atomic Phys. (LPMPS), Univ. Constantine I, Constantine, Algeria \\
$^{af}$ Department of Physics, Faculty of Sciences, University of M'sila, M'sila, Algeria \\
$^{ag}$ Research Unit on Optics and Photonics, UROP-CDTA, S\'etif, Algeria \\
$^{ah}$ Telecom Lab., Faculty of Technology, University Abou Bekr Belkaid, Tlemcen, Algeria \\
$^{ba}$ St. Kliment Ohridski University of Sofia, Bulgaria\\
$^{bb}$ Joint Laboratory of Optics, Faculty of Science, Palack\'{y} University, Olomouc, Czech Republic\\
$^{bc}$ Institute of Physics of the Czech Academy of Sciences, Prague, Czech Republic\\
$^{ca}$ Omega, Ecole Polytechnique, CNRS/IN2P3, Palaiseau, France\\
$^{cb}$ Universit de Paris, CNRS, AstroParticule et Cosmologie, F-75013 Paris, France\\
$^{cc}$ IRAP, Universit\'e de Toulouse, CNRS, Toulouse, France\\
$^{da}$ ECAP, University of Erlangen-Nuremberg, Germany\\
$^{db}$ Karlsruhe Institute of Technology (KIT), Germany\\
$^{dc}$ Experimental Physics Institute, Kepler Center, University of T\"ubingen, Germany\\
$^{dd}$ Institute for Astronomy and Astrophysics, Kepler Center, University of T\"ubingen, Germany\\
$^{de}$ Technical University of Munich, Munich, Germany\\
$^{ea}$ Istituto Nazionale di Fisica Nucleare - Sezione di Bari, Italy\\
$^{eb}$ Universita' degli Studi di Bari Aldo Moro and INFN - Sezione di Bari, Italy\\
$^{ec}$ Dipartimento di Fisica e Astronomia "Ettore Majorana", Universit di Catania, Italy\\
$^{ed}$ Istituto Nazionale di Fisica Nucleare - Sezione di Catania, Italy\\
$^{ee}$ Istituto Nazionale di Fisica Nucleare - Laboratori Nazionali di Frascati, Italy\\
$^{ef}$ Istituto Nazionale di Fisica Nucleare - Sezione di Napoli, Italy\\
$^{eg}$ UniversitaÕ di Napoli Federico II - Dipartimento di Fisica "Ettore Pancini", Italy\\
$^{eh}$ INAF - Istituto di Astrofisica Spaziale e Fisica Cosmica di Palermo, Italy\\
$^{ei}$ Istituto Nazionale di Fisica Nucleare - Sezione di Roma Tor Vergata, Italy\\
$^{ej}$ Universita' di Roma Tor Vergata - Dipartimento di Fisica, Roma, Italy\\
$^{ek}$ Istituto Nazionale di Fisica Nucleare - Sezione di Torino, Italy\\
$^{el}$ Dipartimento di Fisica, Universita' di Torino, Italy\\
$^{em}$ Osservatorio Astrofisico di Torino, Istituto Nazionale di Astrofisica, Italy\\
$^{en}$ Uninettuno University, Rome, Italy\\
$^{fa}$ Chiba University, Chiba, Japan\\ 
$^{fb}$ National Institutes for Quantum and Radiological Science and Technology (QST), Chiba, Japan\\ 
$^{fc}$ Kindai University, Higashi-Osaka, Japan\\ 
$^{fd}$ Kanazawa University, Kanazawa, Japan\\ 
$^{fe}$ Institute for Cosmic Ray Research, University of Tokyo, Kashiwa, Japan\\ 
$^{ff}$ Konan University, Kobe, Japan\\ 
$^{fg}$ Kyoto University, Kyoto, Japan\\ 
$^{fh}$ Shinshu University, Nagano, Japan \\
$^{fi}$ National Astronomical Observatory, Mitaka, Japan\\ 
$^{fj}$ Nagoya University, Nagoya, Japan\\ 
$^{fk}$ Institute for Space-Earth Environmental Research, Nagoya University, Nagoya, Japan\\ 
$^{fl}$ Graduate School of Science, Osaka City University, Japan\\ 
$^{fm}$ Institute of Space and Astronautical Science/JAXA, Sagamihara, Japan\\ 
$^{fn}$ Saitama University, Saitama, Japan\\ 
$^{fo}$ Hokkaido University, Sapporo, Japan \\ 
$^{fp}$ Osaka Electro-Communication University, Neyagawa, Japan\\ 
$^{fq}$ Nihon University Chiyoda, Tokyo, Japan\\ 
$^{fr}$ University of Tokyo, Tokyo, Japan\\ 
$^{fs}$ High Energy Accelerator Research Organization (KEK), Tsukuba, Japan\\ 
$^{ft}$ RIKEN, Wako, Japan\\
$^{ga}$ Korea Astronomy and Space Science Institute (KASI), Daejeon, Republic of Korea\\
$^{gb}$ Sungkyunkwan University, Seoul, Republic of Korea\\
$^{ha}$ Universidad Nacional Aut\'onoma de M\'exico (UNAM), Mexico\\
$^{hb}$ Universidad Michoacana de San Nicolas de Hidalgo (UMSNH), Morelia, Mexico\\
$^{hc}$ Benem\'{e}rita Universidad Aut\'{o}noma de Puebla (BUAP), Mexico\\
$^{hd}$ Universidad Aut\'{o}noma de Chiapas (UNACH), Chiapas, Mexico \\
$^{he}$ Centro Mesoamericano de F\'{i}sica Te\'{o}rica (MCTP), Mexico \\
$^{ia}$ National Centre for Nuclear Research, Lodz, Poland\\
$^{ib}$ Faculty of Physics, University of Warsaw, Poland\\
$^{ja}$ Institute of Space Science ISS, Magurele, Romania\\
$^{ka}$ Joint Institute for Nuclear Research, Dubna, Russia\\
$^{kb}$ Skobeltsyn Institute of Nuclear Physics, Lomonosov Moscow State University, Russia\\
$^{kc}$ Space Regatta Consortium, Korolev, Russia\\
$^{la}$ Institute of Experimental Physics, Kosice, Slovakia\\
$^{lb}$ Technical University Kosice (TUKE), Kosice, Slovakia\\
$^{ma}$ Universidad de Le\'on (ULE), Le\'on, Spain\\
$^{mb}$ Instituto Nacional de T\'ecnica Aeroespacial (INTA), Madrid, Spain\\
$^{mc}$ Universidad de Alcal\'a (UAH), Madrid, Spain\\
$^{md}$ Universidad Polit\'ecnia de madrid (UPM), Madrid, Spain\\
$^{me}$ Instituto de Astrof\'isica de Canarias (IAC), Tenerife, Spain\\
$^{na}$ KTH Royal Institute of Technology, Stockholm, Sweden\\
$^{oa}$ Swiss Center for Electronics and Microtechnology (CSEM), Neuch\^atel, Switzerland\\
$^{ob}$ ISDC Data Centre for Astrophysics, Versoix, Switzerland\\
$^{oc}$ Institute for Atmospheric and Climate Science, ETH Z\"urich, Switzerland\\
$^{pa}$ Space Science Laboratory, University of California, Berkeley, CA, USA\\
$^{pb}$ University of Chicago, IL, USA\\
$^{pc}$ Colorado School of Mines, Golden, CO, USA\\
$^{pd}$ University of Alabama in Huntsville, Huntsville, AL; USA\\
$^{pe}$ Lehman College, City University of New York (CUNY), NY, USA\\
$^{pf}$ NASA Marshall Space Flight Center, Huntsville, AL, USA\\
$^{pg}$ University of Utah, Salt Lake City, UT, USA\\
$^{ph}$ Georgia Institute of Technology, USA\\
$^{pi}$ University of Iowa, Iowa City, IA, USA\\
$^{pj}$ NASA Goddard Space Flight Center, Greenbelt, MD, USA\\
$^{pk}$ Center for Space Science \& Technology, University of Maryland, Baltimore County, Baltimore, MD, USA\\
$^{pl}$ Department of Astronomy, University of Maryland, College Park, MD, USA\\
$^{pm}$ Fairfield University, Fairfield, CT, USA
}
\end{document}

%% file: intro.tex
\section{Introduction}
With a record energy of more than 100~EeV, Ultra High Energy Cosmic Rays (UHECRs) are the most energetic particles known to exist. Although ground-based observatories have observed these energetic particles for decades, their source and acceleration mechanisms remain largely unknown due to their extremely low flux at Earth's surface (see \cite{UHECRoverview} and references therein).

Very High Energy (VHE) neutrinos (E>10~PeV) can also help to explain the most energetic processes in the universe as well as the evolution of astrophysical sources. Few of these VHE neutrinos have been detected so far due to their minuscule interaction cross sections, requiring very large amounts of target material to allow for observation.

A space-based experiment such as the proposed Probe for Multi Messenger Astrophysics (POEMMA) \cite{POEMMA-JCAP} could overcome the shortcomings of ground-based observation (by observing the Earth and its atmosphere from above, allowing for gargantuan increase in acceptance at the highest energies), thereby representing the next frontier in UHECR and VHE neutrino physics. Before such a detector can be built and launched, it is highly advantageous to develop pathfinder missions to raise the technological readiness and to verify the targeted detection techniques. A stratospheric balloon allows for such an investigation in a near space environment without the risk and cost of a fully realized space mission.

\begin{figure}[h!]
    \centering
    \includegraphics[width=0.75\textwidth]{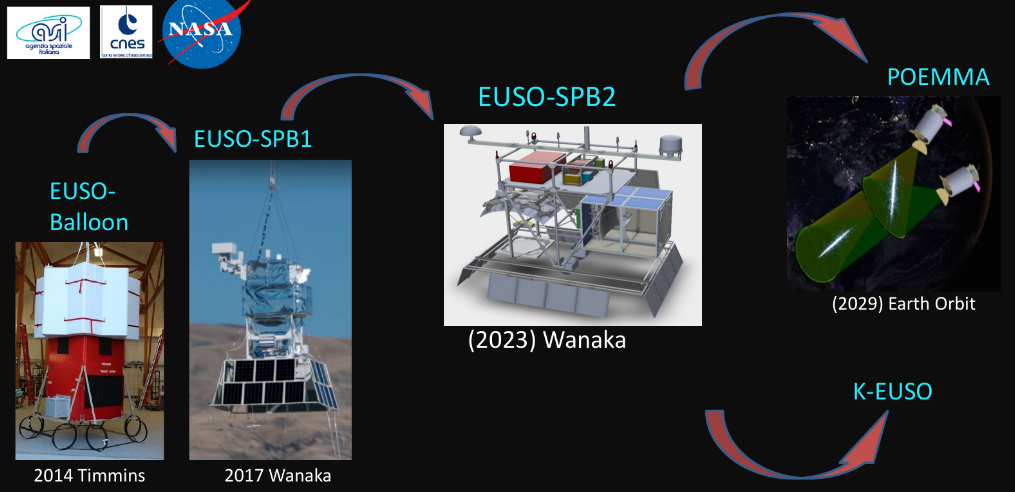}
    \caption{Evolution of the EUSO ballooning program towards a space based mission, starting from the first proof of concept of the technology with EUSO-Balloon in 2014 and updated payload in EUSO-SPB1 in 2017. EUSO-SPB2 will launch in 2023 and is the next step towards a space based mission like K-EUSO or POEMMA.}
    \label{fig:EUSOTimeline}
\end{figure}

The Extreme Universe Space Observatory on a Super Pressure Balloon 2 (EUSO-SPB2) is the third (and most advanced) balloon mission undertaken by the EUSO collaboration and will build on the experiences of previous missions \cite{EUSOBalloon, EUSOSPB1}. The timeline of the evolution of the EUSO ballooning missions is shown in Fig. \ref{fig:EUSOTimeline}.

EUSO-SPB2 has three main scientific objectives:
\begin{enumerate}
    \item Observing the first extensive air showers via the fluorescence technique from suborbital space.
    \item Observing Cherenkov light from upwards going extensive air showers initiated by cosmic rays.
    \item Measuring the background conditions for the detection of neutrino induced upwards going air showers.
    \item Searching for neutrinos from astrophysical transient events (e.g. binary neutron star mergers)
\end{enumerate}
The mission and the instrument will be detailed in section \ref{sec:MissionInstrument} before we discuss the different science objectives for the two telescopes in sections \ref{sec:FTscience} and \ref{sec:CTScience}, respectively.

%% file: Mission_Instrument.tex
\section{The EUSO-SPB2 Instrument and Mission}
\label{sec:MissionInstrument}
To achieve the science goals outlined earlier, EUSO-SPB2 is equipped with two telescopes: one optimized for the fluorescence detection technique (nadir pointing), and a second one optimized for the Cherenkov detection technique (pointing near the Earth limb).

The \textit{Fluorescence Telescope (FT)} camera has a modular design following the layout of previous instruments, particularly EUSO-SPB1, which flew in 2017. While EUSO-SPB1 was equipped with only one Photo Detection Modules (PDM), each consisting of 2304 individual pixels capable of single photo electron counting with a double pulse resolution of 6~ns, EUSO-SPB2 will fly 3 PDMs. Also the integration time was shortened from \unit[2.5]{$\mu$s} to \unit[1]{$\mu$s} in EUSO-SPB2 to increase the signal to noise ratio. The optical system used in EUSO-SPB1 was a 2 Fresnel lens system. To increase the light collection onto the focal surface EUSO-SPB2 design was switched to a Schmidt system consisting of 6 mirror segments and a 1~m diameter Schmidt corrector plate at the aperture providing a field of view of roughly 12$^\circ$ by 36$^\circ$. A detailed description of the FT is given in \cite{SPB2-FT}.

The \textit{Cherenkov Telescope (CT)} is a brand new instrument. To accommodate the very fast signals of the Cherenkov emission, the camera is composed of 512 Silicon Photomultipliers (SiPMs) pixels with a 10ns integration time and a spectral range between 400~nm and 800~nm.
The optical system of the CT is a Schmidt system similar to the FT but with only 4 mirror segments. The field of view of the CT is 6.4$^\circ$ in zenith and 12.8 $^\circ$ in azimuth. The mirror segments are aligned in such a way that a parallel light pulse from outside the telescope produces two spots in the camera. This bi-focal alignment allows to distinguish triggers of interest from single spot events caused by low energy cosmic rays striking the SiPM camera directly. A detailed description of the CT is given in \cite{SPB2-CT}.

EUSO-SPB2 is designed as a payload of a NASA Super Pressure Balloon, allowing a mission duration of up to 100 days at a nominal altitude of 33~km. EUSO-SPB2's planned launch is in March/April of 2023 from Wanaka, New Zealand. This launch window and location is chosen to allow the balloon to travel easterly around the globe following a stratospheric air circulation that develops twice a year in these latitudes ($\sim 45^{\circ}$S). The azimuth rotator will allow to point the solar panels during the day towards the sun for maximum battery charging time for the night-time data taking. The average observation time per night is around 5 hours over the entire flight. While the FT points nadir during the entire flight for the observation of EAS via the fluorescence technique, the CT has a pointing mechanism allowing to aim the telescope both above and below the Earth limb. Above the limb, the CT will record direct Cherenkov light from PeV cosmic rays early in the flight to verify the functionality of the instrument. Below the limb, the CT will search for signals corresponding to the upward $\tau$-induced EAS. In case that an astrophysical event alert is issued during the mission, the zenith and azimuth pointing capability of the CT will allow for a follow up search (see \ref{subsec:ToO}).

%% file: FT_science.tex
\section{Science Objective of the Fluorescence Telescope}
\label{sec:FTscience}
The primary scientific goal of the FT is to achieve the first measurement of an EAS induced by a primary cosmic ray via the fluorescence technique from suborbital space as well as the accurate reconstruction of the properties of such an event (direction and energy). Such a measurement would prove the feasibility of measuring UHECR from space using fluorescence emission--a technique proposed also for the POEMMA mission. To evaluate the capability of the FT to obtain this goal, extensive simulation work was performed. A detailed discussion of the FT performance can be found in \cite{FT-performance}.

\begin{figure}[h]
    \centering
    \includegraphics[width=.6\textwidth]{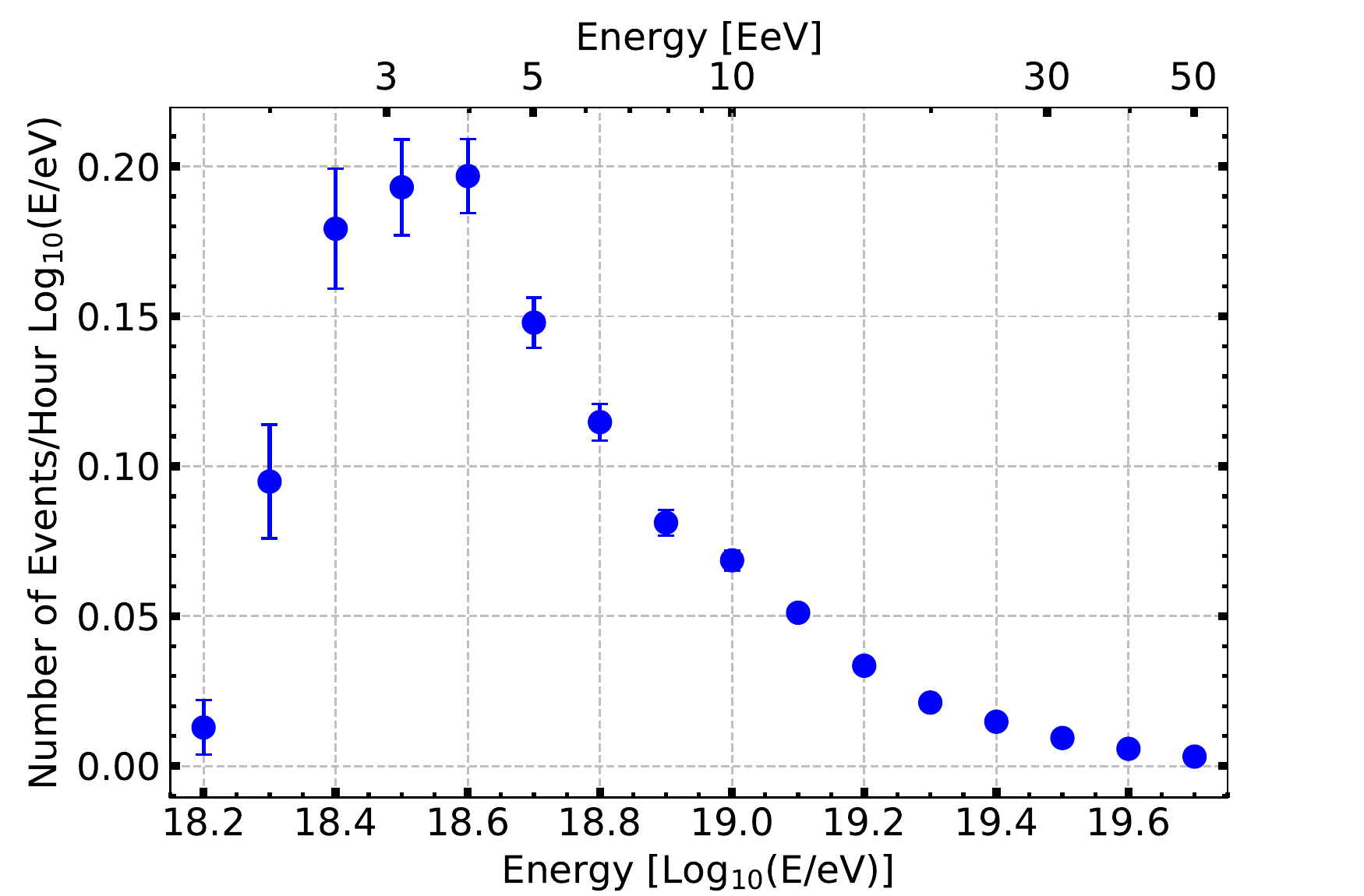}
    \caption{Expected event rate for the FT (See \cite{FT-performance} for more details).}
    \label{fig:FTEventRate}
\end{figure}

For these simulations, CONEX \cite{CONEX} is used to simulate the shower profiles using EPOS-LHC \cite{EPOSLHC} assuming proton primaries. A total of 1.6 million showers were isotropically simulated landing on a disk with \unit[100]{km} radius. The center of the disk is the detector location projected on ground. The 100~km disk is needed to ensure that all shower geometries that deposit light inside the field of view are taken into account. The shower azimuth angle is uniformly distributed between $0^{\circ}$ and $360^{\circ}$ while the zenith angle is distributed flat in $\mbox{sin}\theta\mbox{cos}\theta$ ($0^{\circ}$-80$^\circ$). The energy of the simulated showers is sampled between $10^{17.8}$~eV and $10^{19.7}$~eV in 20 bins evenly spaced in $\mathrm{log}_{10}(E/\mathrm{eV})$. The background used in the simulation is estimated using the measurements from previous experiments appropriately scaled to the new detector characteristics. The Pierre Auger Observatory \cite{Auger_energy} energy spectrum is used to convert a trigger rate to a hourly event rate as shown in fig. \ref{fig:FTEventRate}. 

Integrating over the entire energy span, we expect to record $0.12\pm0.01$ events per observation hour which equates to roughly 0.6 events per night. Machine learning techniques will be used on board to identify these events and download them with the highest priority. This is necessary as a recovery of the whole data set is not guaranteed for a stratospheric balloon flight. More information is provided in \cite{FT-performance}.

Another interesting science objective for the FT concerns the search for the candidate upwards going events as reported by the Antarctic Impulse Transient Antenna (ANITA) \cite{ANITA_upwards}. The feasibility of such a search is still under investigation but preliminary simulations look promising, mostly due to the apparent strong intensity of those events.

%% file: CT_science.tex
\section{Science objective of the Cherenkov telescope}
\label{sec:CTScience}
The CT is a novel instrument and no previous version was flown or even built before. Therefore, the main objective for the telescope is to raise the technical readiness level and to understand the occurring backgrounds in such an instrument before utilizing it in future space missions such as POEMMA. At any given time, the science goals of the CT depend on its pointing direction.
If the instrument is pointed below the limb, there are two objectives that EUSO-SPB2 aims to achieve: i) in principle, detect upwards going EAS sourced from neutrinos from transient astrophysical events (see \ref{subsec:ToO}) and ii) measure optical background signals for the observation of Earth-skimming neutrino events.\\
The background observations are particularly crucial for future missions as currently no measurements in this wavelength range or with this time resolution have been conducted from suborbital space. The night sky background has a significant impact on the detection threshold and the subsequent event reconstruction, which indicates that a detailed measurement of its variation over time and telescope pointing direction over the spectral response of the SiPMs is crucial for the event rate estimation of future experiments. In addition, EUSO-SPB2 will be able to identify known and unknown background sources that could influence the detection of the signal of upwards going EASs for example atmospheric events. Finally, the impact of direct hits on the focal surface cased by charged particles will be studied as they produce the same signature as Cherenkov light from upwards going EASs. EUSO-SPB2 will use the bi-focal technique to identify such events correctly.

\subsection{Target of Opportunity}
\label{subsec:ToO}
\begin{figure}
    \centering
    \includegraphics[width=.9\textwidth]{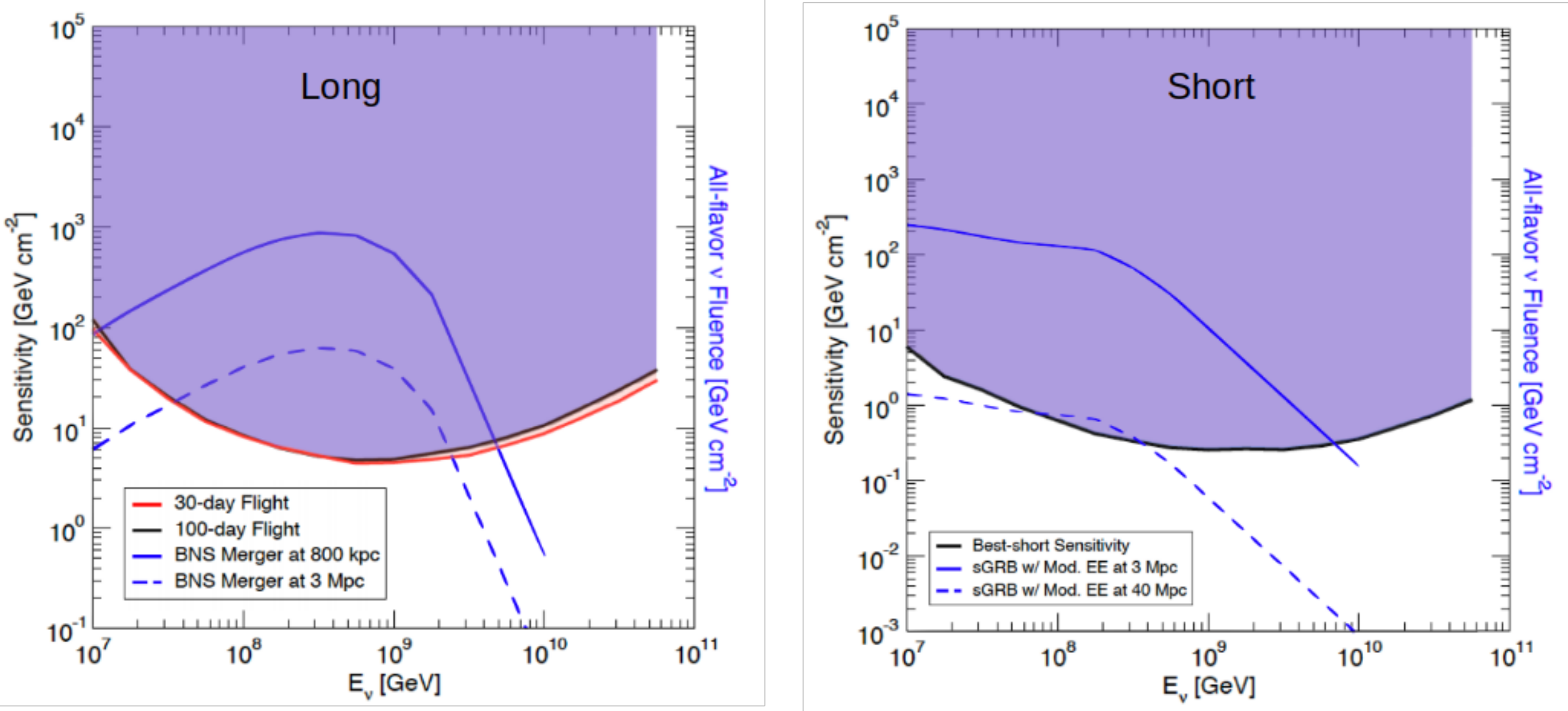}
    \caption{The EUSO-SPB2 all-flavor 90\% unified confidence level sensitivity to ToO neutrino flux per decade in energy. In blue the all-flavor fluency for different models, \cite{Fang_BNNMerger} for the long and \cite{KMMK_sGRB} for the short scenario. For the long duration events a 30 and 100 days flight is considered (red and black line respectively).}
    \label{fig:ToOSensitivity}
\end{figure}

The sensitivity of EUSO-SPB2 to the diffuse cosmogenic neutrino flux was investigated but due to the limited field of view and relatively short mission time, EUSO-SPB2 is not sensitive to the diffuse neutrino flux even assuming the most optimistic UHECR source evolution (corresponding to the cosmological evolution of Active Galactic Nuclei), which is already excluded by the IceCube and Auger experiments. For more details how the sensitivity was estimated, see \cite{NeutrinoSensitivity}. One possibility for EUSO-SPB2 to still observe neutrinos is to point the instrument in the direction of a transient astrophysical event (e.g. binary neutron star mergers after receiving an alert). The EUSO-SPB2 sensitivity to such events, called "target of opportunity" (ToO) events, was investigated in detail in \cite{ToO_poemma}.\\
The ToO events can be divided in two classes based on their characteristic time duration. Short bursts are defined as events which last around 1000~s while long bursts can last from days to even weeks. For the latter, the neutrino flux is averaged over the duration of the event while for the former, it is assumed that the instrument is positioned at the ideal location for the source at the beginning of the event and the flux is integrated over the duration of the event.\\ 
The left panel of fig. \ref{fig:ToOSensitivity} shows EUSO-SPB2's sensitivity for Binary Neutron star mergers \cite{Fang_BNNMerger} scaled for a distance of 0.8~Mpc and 3~Mpc. Possible sources that may correspond to these distances could be located in M31, and NGC 253 and M82 (star burst) respectively. The red line shows a flight duration of 30 days while the black line is for 100 days. The right panel shows the sensitivity for emissions from short gamma-ray burst assuming a moderate emission as discussed in \cite{KMMK_sGRB}. The distances chosen are 3~Mpc and 40~Mpc which correspond to the distances of GW170817 and GRB170817A respectively.
The sensitivities shown in fig. \ref{fig:ToOSensitivity} are not taking the balloon trajectory  into account. For the long scenario the effect of the sun and moon is considered including a longer twilight period due to the balloon altitude. The short scenario does not contain these restrains. Further, although EUSO-SPB2 is sensitive to neutrinos produced in these events, the short mission duration of up to 100 days makes the probability of such an event occurring in close enough range during operations unlikely.

\subsection{Above-the-limb Extensive Air Showers}
\label{subsec:AboveLimb}
\begin{figure}[ht]
    \centering
    \includegraphics[width=.9\textwidth]{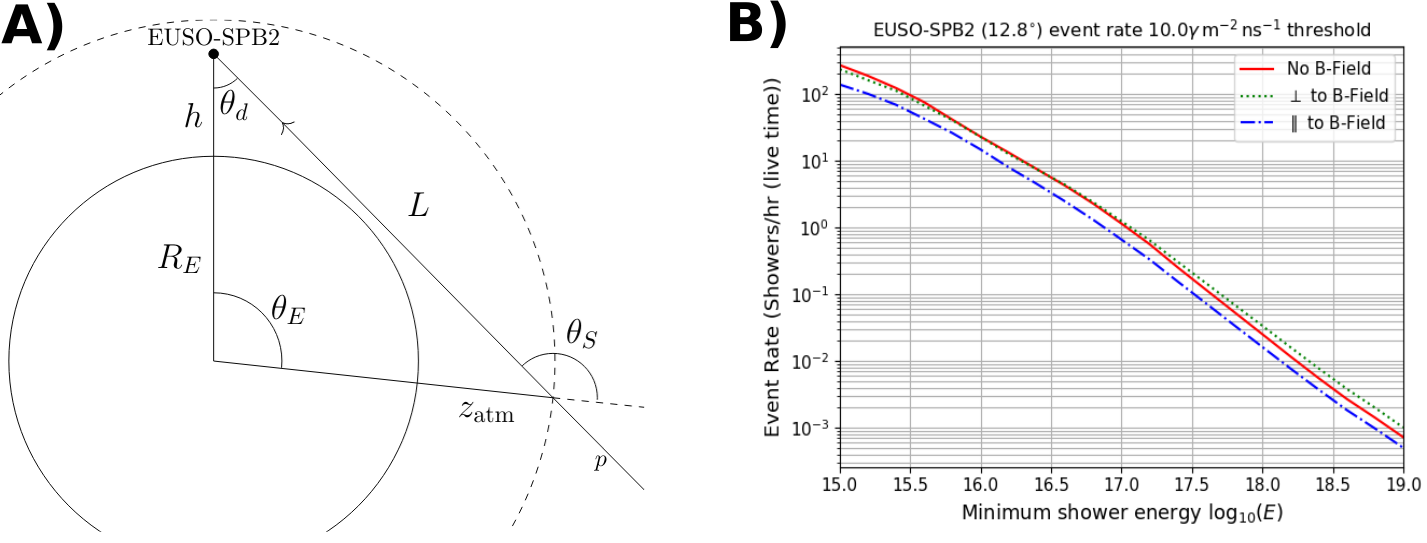}
    \caption{A) Schematic trajectory of upwards going EAS initiated by cosmic rays, observable at the detector via Cherenkov emission of the shower B) The cumulative, energy dependent, expected event rate of above-the-limb cosmic rays, assuming a detection threshold of 10~photons~m$^{-2}$ns$^{-1}$ (See \cite{aboveLimb} for details). The cosmic ray fluxes are taken from \cite{Ralf_book}.}
    \label{fig:AboveLimb}
\end{figure}

When the instrument is pointed above the limb, EUSO-SPB2 will be able to, for the first time, measure Cherenkov emission from EAS from suborbital space. The trajectory of the above-the-limb cosmic ray events is shown in panel A of fig. \ref{fig:AboveLimb}. The trajectories of these events provide enough atmosphere for a primary cosmic ray to produce an EAS which is detectable via its Cherenkov emission as long as the primary energy is larger than a few PeV. This is due to the highly forward beam nature of Cherenkov radiation. As the shower mostly develops in rarefied atmosphere, the atmospheric attenuation is minimized but the Cherenkov emission is also limited due to increased thresholds. Details of how to accurately simulate the Cherenkov signal of such showers is discussed in \cite{aboveLimb}.

The resulting event rate of the above-the-limb cosmic rays for EUSO-SPB2 is displayed in panel B of fig. \ref{fig:AboveLimb}. The cumulative event rate is more than 100 events per hour for primary energies above 1~PeV. Such a high rate makes it possible to collect a statistically relevant dataset in a relatively short observation time, thereby providing a guaranteed signal which has similar properties to those signals from air showers sourced from Earth-skimming neutrinos. Frequent observation of such events will help evaluate and refine the detection and reconstruction techniques of the experiment.

%% file: Conclusion.tex
\section{Conclusion}
EUSO-SPB2 is the follow up project of EUSO-SPB1 and represents a pathfinder for the POEMMA mission and, as such, is equipped with two telescopes. One telescope is optimized for the fluorescence detection of UHECRs while the other is optimized to measure the Cherenkov emission from EAS sourced from either Earth skimming neutrinos or above-the-limb cosmic rays. EUSO-SPB2 is scheduled to fly as a NASA SPB payload from Wanaka, NZ in early 2023.

EUSO-SPB2 will measure, for the first time, EAS from above using the fluorescence technique. Extensive simulation studies have shown that the expected event rate is 0.12 UHECR events per hour with the FT and the possibility to search for ANITA event-like candidates are still under study.
For the first time, a Cherenkov Telescope based on SiPMs will be flown on a stratospheric balloon raising the technological readiness level. EUSO-SPB2 will study the background signals encountered while looking for signals from upward going EASs sourced from below the limb. Although studies have revealed that EUSO-SPB2 has very limited sensitivity towards the diffuse cosmogenic neutrino flux, being significantly less competitive than existing ground based experiments, the instrument still has the capability to detect neutrinos emitted by astrophysical events if they happen within our cosmic neighbourhood. Pointing the CT above the limb allows to record more than 100 events per hour of above-the-limb direct cosmic rays. As this signal is similar to that from potential neutrino induced showers, these events allow to evaluate the detection technique.

\footnotesize{\textit{Acknowledgment:} 
This work was partially supported by Basic Science Interdisciplinary Research Projects of 
RIKEN and JSPS KAKENHI Grant (22340063, 23340081, and 24244042), by 
the Italian Ministry of Foreign Affairs	and International Cooperation, 
by the Italian Space Agency through the ASI INFN agreements n. 2017-8-H.0 and n. 2021-8-HH.0,
by NASA award 11-APRA-0058, 16-APROBES16-0023, 17-APRA17-0066, NNX17AJ82G, NNX13AH54G, 80NSSC18K0246, 80NSSC18K0473, 80NSSC19K0626, and 80NSSC18K0464 in the USA,  
by the French space agency CNES, 
by the Deutsches Zentrum f\"ur Luft- und Raumfahrt,
the Helmholtz Alliance for Astroparticle Physics funded by the Initiative and Networking Fund 
of the Helmholtz Association (Germany), 
by Slovak Academy of Sciences MVTS JEM-EUSO, by National Science Centre in Poland grants 2017/27/B/ST9/02162 and
2020/37/B/ST9/01821, 
by Deutsche Forschungsgemeinschaft (DFG, German Research Foundation) under Germany's Excellence Strategy - EXC-2094-390783311,
by Mexican funding agencies PAPIIT-UNAM, CONACyT and the Mexican Space Agency (AEM), 
as well as VEGA grant agency project 2/0132/17, by the MEYS project CZ.02.1.01/0.0/0.0/17\_049/0008422, and by State Space Corporation ROSCOSMOS and the Interdisciplinary Scientific and Educational School of Moscow University "Fundamental and Applied Space Research".
}